\documentclass[english,groupedaddress, superscriptaddress,aps,10pt, nofootinbib, notitlepage]{revtex4-2}
\usepackage{mathtools}
\usepackage{amsmath}
\usepackage{amssymb}
\usepackage{bigints}
\usepackage{graphicx}
\usepackage{gensymb}
\usepackage{float}
\usepackage[normalem]{ulem}
\usepackage{comment}
\usepackage{siunitx}
\DeclareSIUnit\fps{fps}

\makeatletter

\renewcommand*{\p@subsection}{}

\renewcommand*{\p@subsubsection}{}  

\makeatother
\usepackage{geometry}
\usepackage{bigints}
\usepackage{graphicx}
\usepackage{gensymb}
\usepackage{mathtools}
\usepackage{amsmath}
\usepackage{amssymb}
\usepackage[T1]{fontenc}
\usepackage[utf8]{inputenc}

\DeclarePairedDelimiter\abs{\lvert}{\rvert}%
\DeclarePairedDelimiter\norm{\lVert}{\rVert}%

\usepackage{color}
\definecolor{blue}{rgb}{0,0,1}
\definecolor{black}{rgb}{0,0,0}
\newcommand{\revise}[1]{\textcolor{black}{#1}}
\definecolor{dgreen}{rgb}{0,0.5,0}
\definecolor{dred}{rgb}{0.5,0,0}
\definecolor{dyellow}{rgb}{0.75,0.75,0}

\makeatletter
\let\oldabs\abs
\def\abs{\@ifstar{\oldabs}{\oldabs*}}
\let\oldnorm\norm
\def\norm{\@ifstar{\oldnorm}{\oldnorm*}}
\makeatother

\makeatletter
\usepackage{enumerate}%
\usepackage[caption=false]{subfig}

\begin{document}
\title{Predicting the power grid frequency of European islands}

\author{Thorbjørn Lund Onsaker}
\affiliation{Faculty of Science and Technology, Norwegian University of Life Sciences, 1432 Ås, Norway}

\author{Heidi S. Nygård}
\affiliation{Faculty of Science and Technology, Norwegian University of Life Sciences, 1432 Ås, Norway}

\author{Damià Gomila}
\affiliation{Instituto de Física Interdisciplinar y Sistemas Complejos, IFISC (CSIC-UIB), Campus Universitat Illes Balears, E-07122 Palma de Mallorca, Spain}

\author{Pere Colet}
\affiliation{Instituto de Física Interdisciplinar y Sistemas Complejos, IFISC (CSIC-UIB), Campus Universitat Illes Balears, E-07122 Palma de Mallorca, Spain}

\author{Ralf~Mikut}
\affiliation{Institute for Automation and Applied Informatics, Karlsruhe Institute of Technology, 76344 Eggenstein-Leopoldshafen, Germany}

\author{Richard~Jumar}
\affiliation{Institute for Automation and Applied Informatics, Karlsruhe Institute of Technology, 76344 Eggenstein-Leopoldshafen, Germany}

\author{Heiko Maass}
\affiliation{Institute for Automation and Applied Informatics, Karlsruhe Institute of Technology, 76344 Eggenstein-Leopoldshafen, Germany}

\author{Uwe~Kühnapfel}
\affiliation{Institute for Automation and Applied Informatics, Karlsruhe Institute of Technology, 76344 Eggenstein-Leopoldshafen, Germany}

\author{Veit~Hagenmeyer}
\affiliation{Institute for Automation and Applied Informatics, Karlsruhe Institute of Technology, 76344 Eggenstein-Leopoldshafen, Germany}


\author{Benjamin Schäfer}
\affiliation{Institute for Automation and Applied Informatics, Karlsruhe Institute of Technology, 76344 Eggenstein-Leopoldshafen, Germany}

\begin{abstract}
Modelling, forecasting and overall understanding of the dynamics of the power grid and its frequency are essential for the safe operation of existing and future power grids. Much previous research was focused on large continental areas, while small systems, such as islands are less well-studied. These natural island systems are ideal testing environments for microgrid proposals and artificially islanded grid operation. 
In the present paper, we utilize measurements of the power grid frequency obtained in European islands:  the Faroe Islands, Ireland, the Balearic Islands and Iceland and investigate how their frequency can be predicted, compared to the Nordic power system, acting as a reference. 
The Balearic islands are found to be particularly deterministic and easy to predict in contrast to hard-to-predict Iceland. Furthermore, we show that typically 2-4 weeks of data are needed to improve prediction performance beyond simple benchmarks.
\end{abstract}

\maketitle
\makeatother

\section{Introduction}

An increasingly interdisciplinary community is contributing to the understanding of power grid dynamics and stability, including dynamical modelling, simulations and machine-learning-based forecasts, see e.g. \cite{witthaut2022collective} for a recent review.
The energy system is a very important and challenging research area as it constitutes a complex system with many interacting entities, e.g. its energy markets, generators, consumers, different control areas and control mechanisms.
%

In this article, we investigate the dynamics and predictability of the power grid frequency in five grids of very different sizes. The grid frequency is critical to ensure a safe and reliable supply of electricity due to its central role in power system control \cite{stability}. 
The electric frequency of the alternating current power grids mirrors the current balance in supply and demand: An excess of demand effectively draws from the kinetic energy stored in rotating turbines and slows these down. Vice versa, an excess of generation cannot be stored directly in the grid and thereby leads to an increased frequency (unless batteries, pumped hydropower or other storage options are set in operation to restore the power balance). 
Critically, the frequency has to be controlled tightly using numerous control mechanisms \cite{stability,reserves,hvdc_reserves}, as otherwise consumers or generators have to be disconnected. Desired frequency ranges depend on the synchronous area, i.e. the connected power grid sharing the same frequency, and range from $f=50\pm0.2$ Hz \cite{Iceland_freq} to $f=50\pm0.05$ Hz \cite{ce_frequency} for the areas considered in the present paper. 
With an increasing share of volatile renewable generation, in particular wind and solar power, the total inertia of the power system decreases, while fluctuations overall increase \cite{ulbig2014impact,milano2018foundations}. 
This further amplifies the need for accurate forecasts and a solid understanding of power systems. 

Islands and their power grids are particularly interesting for two reasons: 
First, there exist many natural islands which are unconnected to any other power grid or will only be connected via HVDC (high-voltage-directed-current) links, still ensuring decoupled frequency dynamics \cite{pillai2010vehicle}. 
Small island grids typically have lower inertia and display larger frequency deviations than large integrated networks \cite{gorjao2020data}. Hence they may provide valuable insights for variable grid operation.
Second, microgrid proposals \cite{parhizi2015state,microgrids,microgrids2}, which suggest establishing a local cell of the power grid, are receiving much attention. Depending on the specific scenario, cells might be mostly autonomous, have their own market \cite{palizban2014microgrids} or only decouple during emergencies to stabilize the grid and avoid cascading failures \cite{teymouri2017toward,schafer2018dynamically}.

Fully modelling the short-term frequency dynamics is very challenging due to its combined stochastic-deterministic nature and the numerous external influences \cite{schafer2018non,schafer2018isolating,gorjao2020data}, suggesting the use of data-driven approaches \cite{vorobev2019deadbands} and the usage of machine learning to predict time series as well as to understand and control energy systems \cite{Benjamin, hossain2019application,yang2021power,kruse2021revealing}. 
These data-driven approaches complement modelling-based approaches \cite{gorjao2020data,kyesswa2020dynamic} as they do not make any assumptions about the governing dynamical equations but still provide forecasts and explanations of the system.

Any data-based forecasting method depends on the availability of data with a sufficient quality covering a sufficiently long period.
Unfortunately, such data are still not easily available. Even research projects, such as the GridEye/FNET initiative \cite{chai_wide-area_2016}, are not always open, which limits their value for the research community.
In  recent work, we have pushed for more open data of power grid frequency measurements \cite{benjamin_2,jumar2020database,power_grid_frequency}, including coverage of  power grids on islands. Still, with only limited measurement devices available, some questions arise: Are previously developed methods which forecast the power grid frequency \cite{Benjamin} applicable to islands? How much data are necessary to develop data-driven approaches that can outperform simple benchmarks? How can additional information, e.g. from transparency platforms \cite{ENTSOETransparencyPlatform}, be integrated into frequency forecasts?

To answer these questions, the remaining paper is structured as follows. We start with an overview of the data obtained from the Nordic system, the Faroe Islands, Ireland, the Balearic Islands and Iceland in Sec.~\ref{sec:overview}. Next, we provide an introduction to the nearest-neighbour predictor used for forecasting in Sec.~\ref{sec:Intro_predictor}. Then, we investigate the performance of the predictor, compared to benchmarks and as a function of available data in Sec.~\ref{sec:perform_predcitor} to then discuss the possibility of extending the predictor by including additional data in Sec.~\ref{sec:additonal_feat}. We close with a discussion in Sec.~\ref{sec:discssion}.

\section{Data overview}\label{sec:overview}

\begin{figure}[h]
    \includegraphics[width=13.5cm]{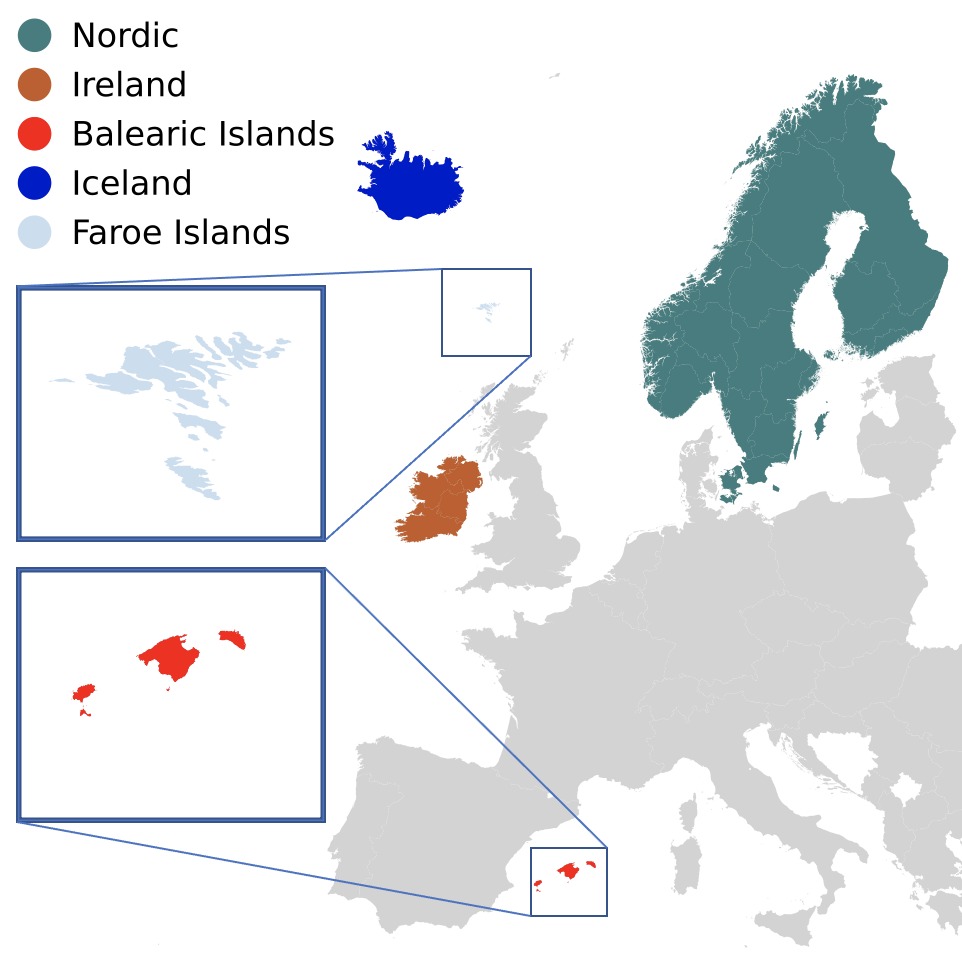} 
    \caption{Frequency data are taken from several European islands. We display the synchronous areas analysed within the present article: Ireland, Balearic Islands, Iceland and the Faroe Islands as the island systems with the Nordic synchronous area as a continental reference system. Map created via Basemap and Python 3.
    \label{fig:map}}
\end{figure}

In the present article, we analyse the power grid frequency time series from four different European islands in addition to the Nordic synchronous area. These islands are the Faroe Islands, Ireland, the Balearic Islands and Iceland, see also Fig~\ref{fig:map}, showing good geographical spread and a difference in size. We consider time series from 6 days (Faroe Islands) up to 450 days (Balearic Islands) for our analysis. 
The five regions (Nordic and the islands) are also very different in terms of their energy mix, regulations and interconnections:

The Nordic power system includes Norway, Sweden and Finland as well as the Eastern part of Denmark, a total of approximately 24 million people. The energy mix for the Nordic electricity supply consists of a significant share of renewable energy, in particular wind and hydropower, with a steady increase in recent years \cite{norden}.

The Faroe Islands are an archipelago of 18 major islands in the Atlantic Ocean, with approximately 50 000 inhabitants. The power generation relies mostly on fossil fuels \cite{faroe}. 

The Irish power system covers both the Republic of Ireland and Northern Ireland, with about 6.8 million inhabitants in total. Ireland's generation is tailed towards a high share of wind energy and in 2020 renewable electricity generation reached 43\% of annual generation \cite{soni}. Ireland is connected to Great Britain via HVDC cables \cite{putz2022revealing}.

The Balearic Islands, belonging to Spain, are located in the Mediterranean Sea with a total population of about 1.2 million. Generation relies mostly on fossil fuels such as gas (combined cycle gas turbines) and diesel engines leading to a share in generation of more than 70\% \cite{mallorca_studie}. Furthermore, an HVDC cable connecting the islands to mainland Spain is an important factor in the electric system.

Finally, Iceland is located in the North Atlantic Ocean and has roughly 300 000 inhabitants. Its main source of electricity generation is hydropower, making up 75\% of the total generation. In addition, Iceland's volcanic activities allow extensive geothermal power generation. These cheap energy sources have attracted energy-intensive industries such as ferrosilicon, aluminium smelters and data centres \cite{iceland}.

The data for the Nordic grid are based on measurements provided by the Finish operator Fingrid \cite{Fingrid}, see also \cite{benjamin_2,jumar2020database,power_grid_frequency}. Meanwhile, the time series of the islands are obtained via our measurement device, the Electrical Data Recorder (EDR), which has been developed at KIT. The EDR records the voltage with a sampling frequency of \SI{25}{kHz} and assigns a standard time to each data point based on a GPS recorder. Then, the frequency is extracted based on zero-crossings of the voltage signal, see ~\cite{maass2013first, maas2015data, jumar2020database, foerstner2022experimental} for details.
Finally, the frequency data are pre-processed to identify acquisition interruptions and to remove outliers. Unfortunately, some single devices had some longer measurement outages, leading to  up to 25\% of missing data related to the installed time period in the Balearic Islands and Ireland, respectively, but typically less than 1\% for the other locations, see also code for details \cite{github}.

\begin{figure}[h]
    \includegraphics[width=13.5cm]{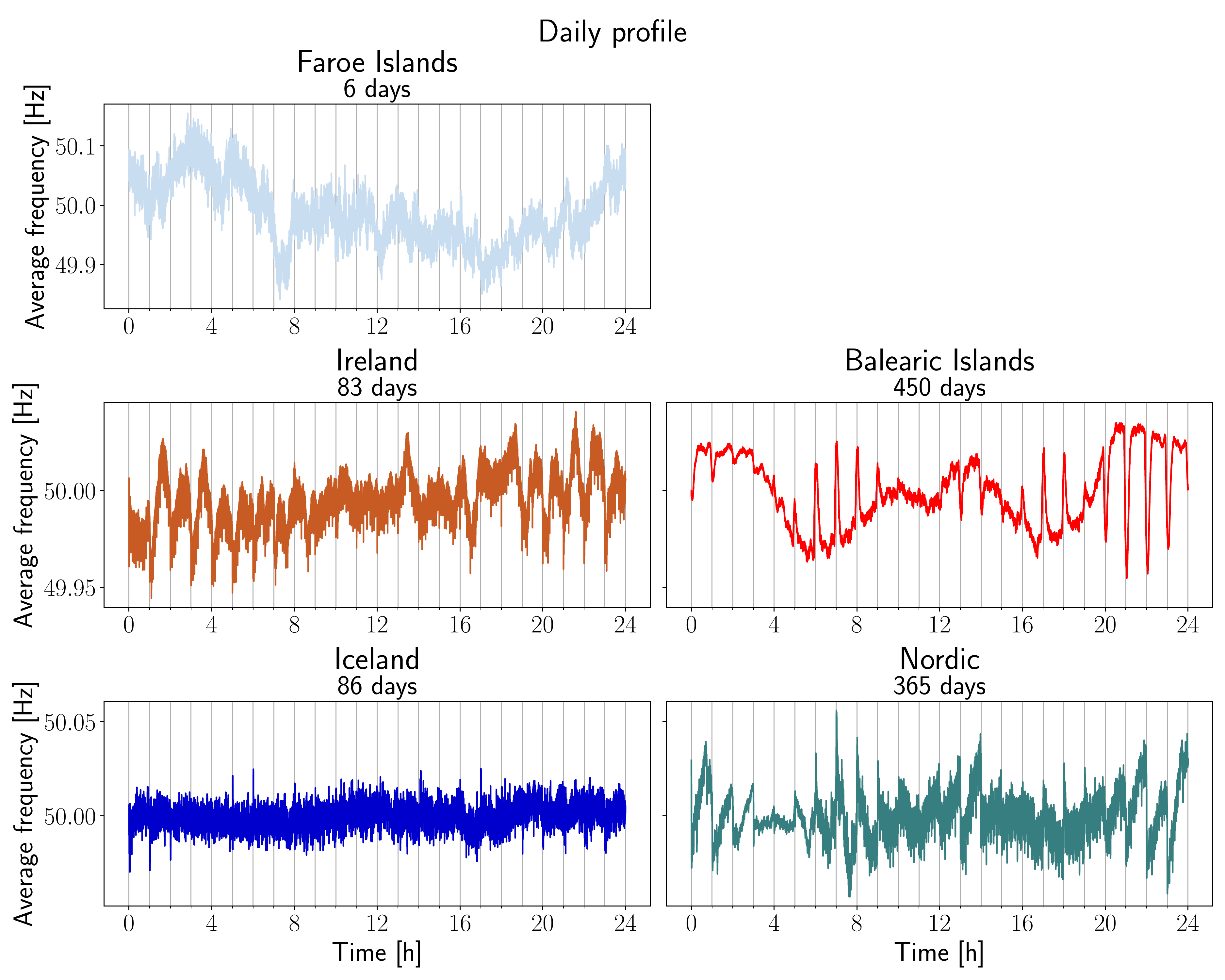} 
    \caption{The daily profile acts as the benchmark for all regions. We define the daily profile as the average of all recordings for every second of the day, averaged over all days in the data set, see also eq.~\eqref{eq:daily_profile}. 
    We note that the daily profiles in the Nordic and Balearic grids are not significantly affected if the time series is truncated to shorter periods, see also the code \cite{github}.
    \label{fig:dailyProfile}}
\end{figure}

To obtain an initial impression of the frequency dynamics in the different regions, we compute the daily profiles in Fig.~\ref{fig:dailyProfile}, noting stark differences.
The daily profile is the average frequency for each second of a day, averaged over all days in the data set, defined in detail in the next section. 
The daily profile in the Faroe Islands displays the largest deviations and seems to be somewhat unpredictable. This is in part due to the short time series available (6 days) and might further be linked to the small population size. Meanwhile, Ireland, the Balearic Islands and the Nordic grid all display clear patterns with the frequency jumping up and down at specific hours. These jumps are associated with typical power dispatch actions and load and generation ramps at full hours \cite{kruse2021exploring, weissbach2009high}. Finally, the Icelandic daily profile is mostly flat and does not display clear patterns, likely because the dominant hydropower and geothermal power plants are easily controllable on the generation side. 

\begin{figure}[h]
    \includegraphics[width=13.5cm]{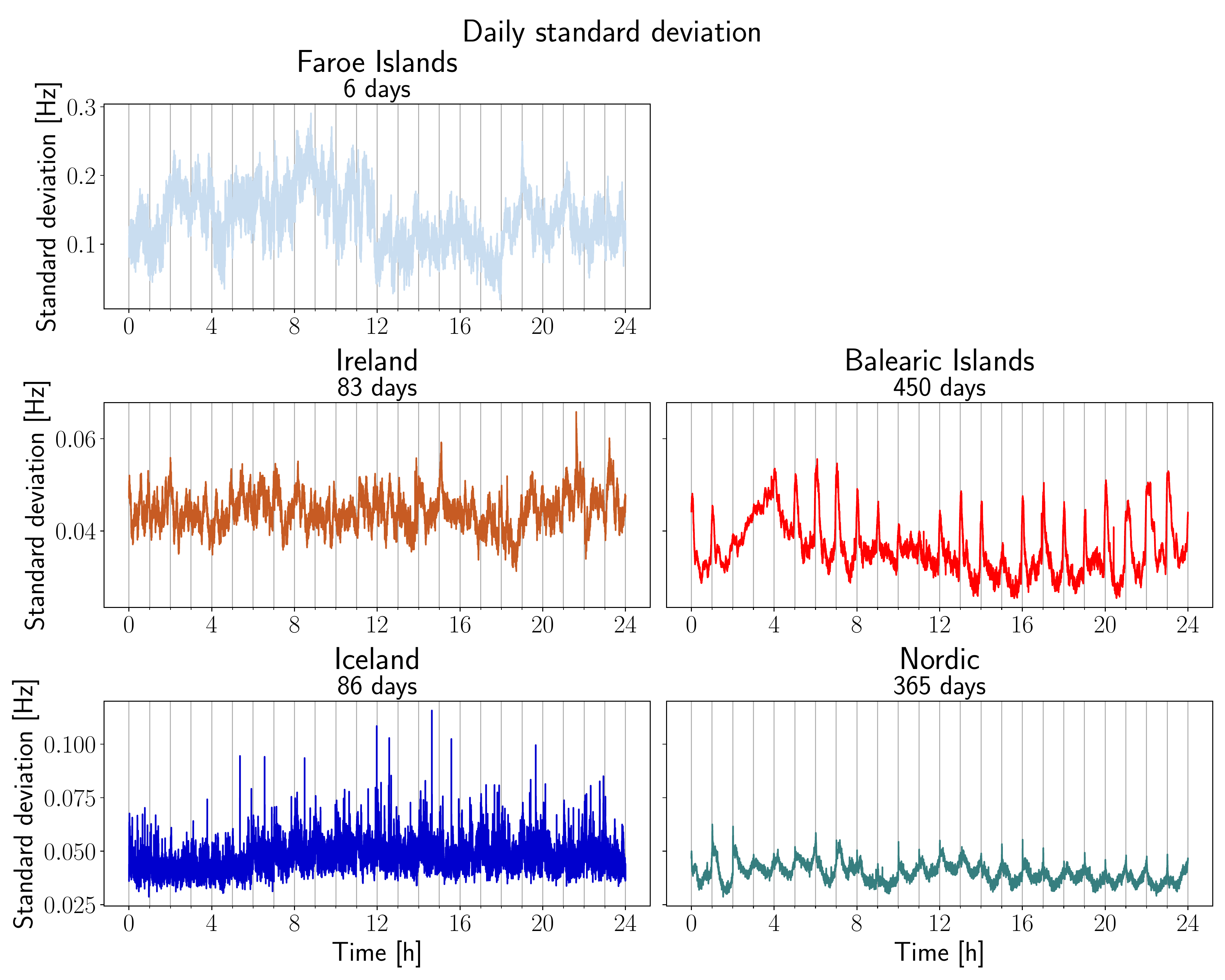} 
    \caption{
    The daily standard deviation reflects the variability within the frequency samples at each second of the day. The Faroe Islands and Iceland show particularly large deviations, in contrast to the deterministic profiles in Ireland, the Balearic Islands and the Nordic grid.
\label{fig:dailySTD}}
\end{figure}

While the daily profile (mean daily frequency values) provides insights into what typical frequency values to expect, the daily standard deviation reveals the spread of frequency values over the day, see Fig.~\ref{fig:dailySTD}. Technically, instead of computing the mean value, we compute the standard deviation for each second of a day.
Analysing the standard deviation confirms a previous observation from the daily profile: The Balearic Islands, the Nordic and to a slightly lesser extent the Irish systems show clean curves with regular distinct peaks at the hourly dispatch intervals. This is consistent with earlier observations: Larger deterministic frequency excursions at the start of an hour are complemented with smaller but random fluctuations in between \cite{weissbach2009high,schafer2018non,gorjao2020data}. 
Meanwhile, the Faroe Islands show little structure and the largest absolute standard deviation, consistent with the large frequency deviations observed earlier. 
Iceland is special in that the standard deviation is on average higher than in Ireland, the Balearic Islands or the Nordic grid and occasionally, very large values of the standard deviations are observed due to extreme frequency deviation of the order of Hertz, instead of deviations of tens or maybe one hundred millihertz as would be observed in the other synchronous areas.

\begin{figure}[h]
    \includegraphics[width=13.5cm]{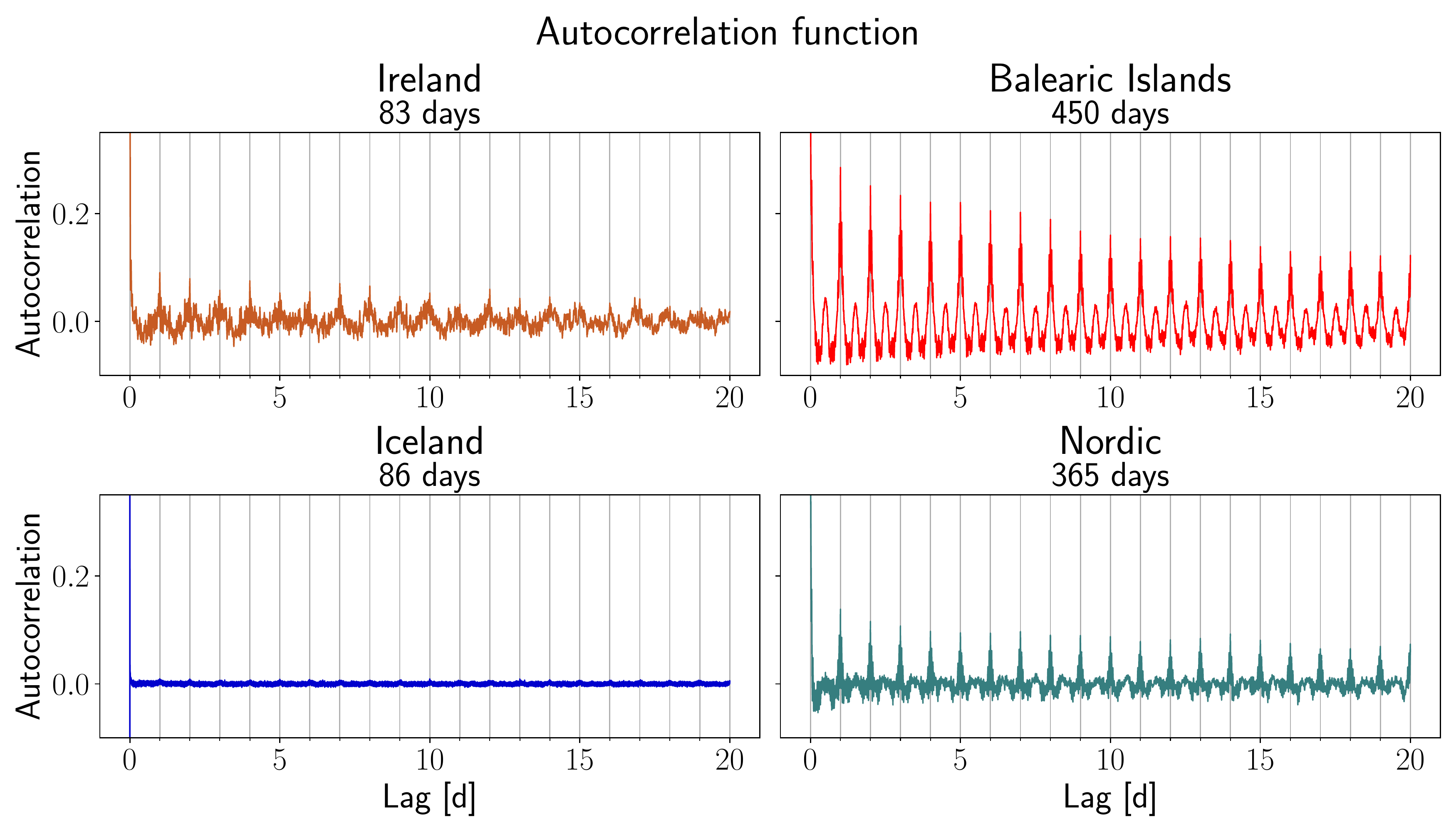} 
    \caption{The autocorrelation reveals daily and other regular patterns in three out of the four island grids.
    We plot the autocorrelation function with time lags up to 20 days and notice very pronounced patterns in the Balearic Islands but almost no patterns in Iceland. 
    \label{fig:autocorrelation}}
\end{figure}

Complementing the daily profile and standard deviation, we analyse how much history is preserved in the trajectories by computing the autocorrelation decay \cite{schafer2018non}. 
For a generic stochastic time series, we might expect an exponential decay of the autocorrelation \cite{gardiner1985handbook} and we do indeed observe such a (quick) decay in all time series, see Fig.~\ref{fig:autocorrelation}. For all grids except Iceland, we also observe regular peaks in the autocorrelation on a daily basis, reinforcing the idea that the frequency follows daily patterns. Iceland is special in that the frequency displays almost no correlation and seems to be much more stochastic than the frequency in the other regions. Note that the Faroe Islands are omitted due to their short time series. 

\begin{figure}[h]
    \includegraphics[width=13.5cm]{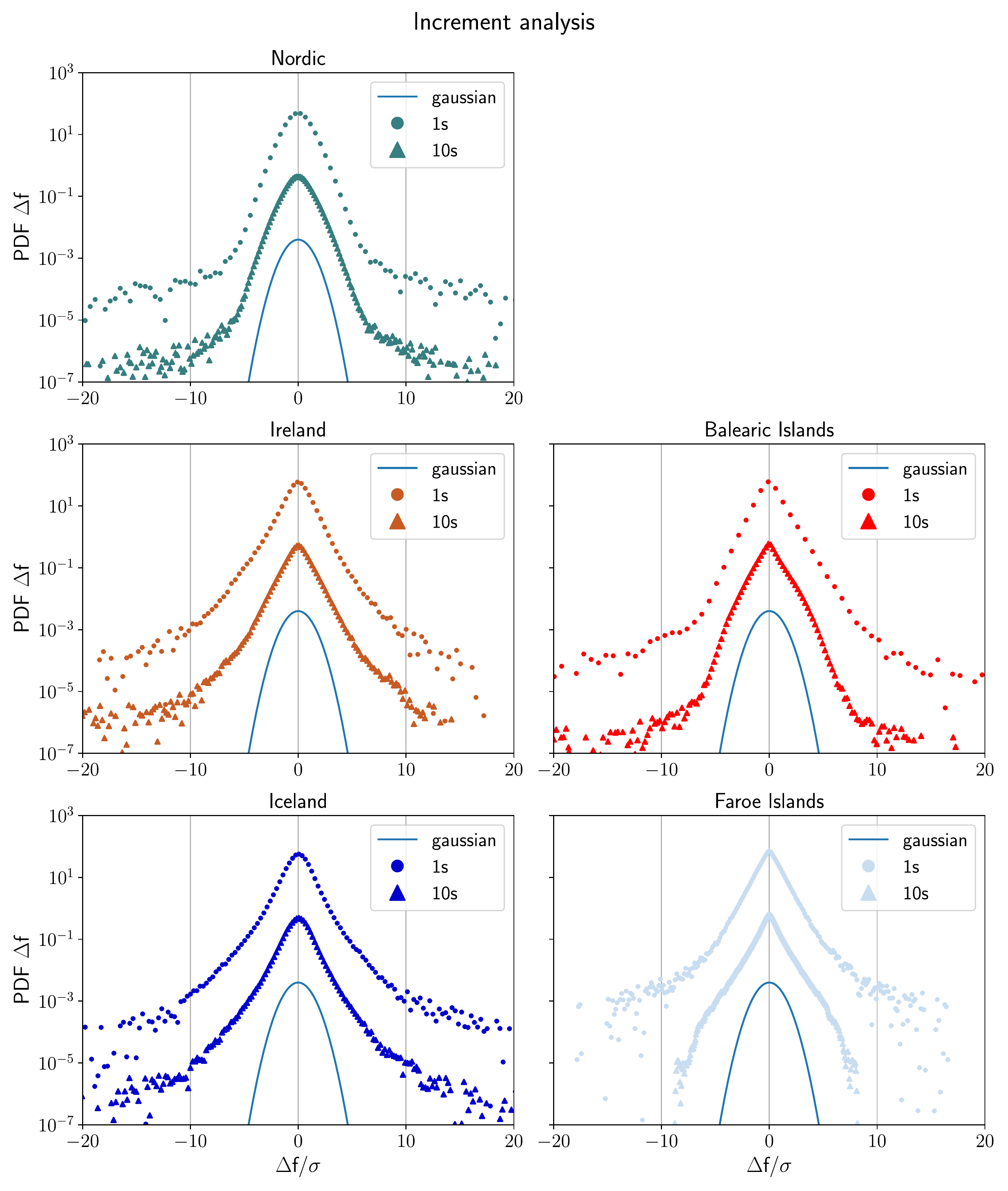} 
    \caption{The increments statistics of all power grids deviates from Gaussian increments, particularly on short time scales. We plot the frequency increments $\Delta f_\tau=f(t+\tau)-f(t)$ for delays of $\tau=1s$ (circles) and $\tau=10s$ (triangles), normalized with the respective standard deviation of a grid $\sigma$ and plot a Gaussian curve (blue solid line) for reference. Curves are shifted vertically for better visibility.
    \label{fig:increments}}
\end{figure}

\revise{Finally, we also investigate the increment distribution, see Fig.~\ref{fig:increments}. The increments are given as $\Delta f_\tau = f(t+\tau)-f(t)$. If the system dynamics followed a simple stochastic process, e.g. an Ornstein-Uhlenbeck \cite{gardiner1985handbook,gorjao2020data}, the increments should resemble a Gaussian distribution on all time scales $\tau$. In practice, we observe that for smaller $\tau$ values, the deviations from Gaussianity are more pronounced, while for larger values of $\tau$, the increments become more Gaussian \cite{benjamin_2, anvari2020stochastic}. We observe the same effect in  Fig.~\ref{fig:increments}: All synchronous areas display deviations from Gaussianity on short time scales ($\tau=1s$, circles), while the statistics approach a more Gaussian shape for longer time scales ($\tau=10s$, triangles). Ireland and even more so Iceland still display strong non-Gaussian distributions for $\tau=10s$.
}

\section{Forecasting frequency on islands}

\subsection{Introducing the WNN predictor}\label{sec:Intro_predictor}

\begin{figure}[h]
    \includegraphics[width=13.5cm]{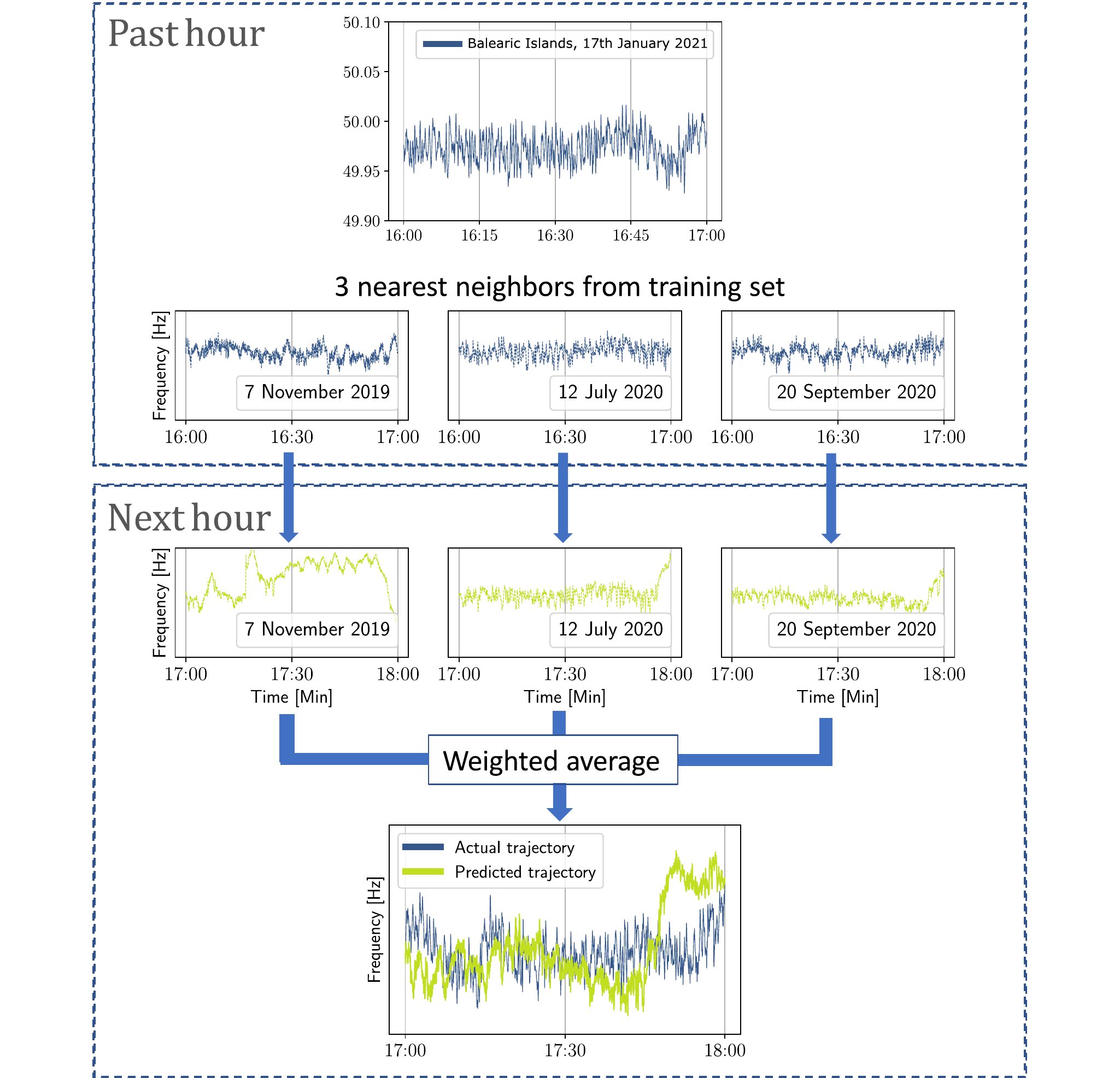} 
    \caption{The WNN predictor predicts the next hour trajectory based on similar patterns from the past. The past hour is compared with patterns from the training set, for which the next hour is available, and patterns most similar to the past hour are identified. Here, as an example, the three nearest neighbours are shown. A weighted average of the nearest neighbours’ next hour trajectory is computed as the prediction and can thus be compared to the actual trajectory from the test set.
    \label{fig:WNN_illustration}}
\end{figure}

To forecast the power grid frequency, we expand upon an earlier paper \cite{Benjamin}, which  investigated numerous aspects of frequency predictability with a weighted nearest neighbour (WNN) predictor. 
This earlier paper \cite{Benjamin}  focused on the three largest synchronous areas in Europe, for which several years of data are available. In contrast, measurements from islands, investigated in the present article, cover a few weeks or months up to 450 days in the case of the Balearic islands.
The WNN algorithm identifies similar patterns in the past to forecast future dynamics, see Fig.~\ref{fig:WNN_illustration} for an illustration.

Let us specify the details of the prediction process:
For each synchronous area, the time series is split into chunks of equal length which are then used to find similar chunks (neighbours) from the past. More formally, we define the pattern vectors
\begin{equation}
\mathbf F_{n} = \begin{pmatrix} f(t_{0} -(n+1)\gamma \tau) \\ f(t_{0} - (n+1)\gamma \tau + \tau)\\ f(t_{0} - (n+1)\gamma \tau + 2\tau) \\ {\dots }\\ f(t_{0} - n\gamma \tau - \tau) \end{pmatrix}, \label{F}
\end{equation} 
where $t_0$ is the last time point for which we have data in our initial pattern $\mathbf F_{0}$. Furthermore, the time delay $\tau$ is identical with the time resolution $\tau=1$s and we choose  $\gamma=3600$ data points so that $\gamma\tau = 3600$s covers one hour, a central time scale for market dynamics and regular grid patterns \cite{weissbach2009high}. 
Hence, each pattern $\mathbf F_{n}$ is a vector with 3600 frequency measurements and the entire time series is covered by combining non-overlapping sets as
\begin{equation} 
\mathcal F = \{ \mathbf F_{n} | \exists i \in \mathbb N: n\gamma \tau = i\cdot 24 \textrm {h} \} \label{time_sensitive}
\end{equation}
and the index $n$ is used to distinguish between different patterns. 
Given the past hour pattern as $\mathbf F_{0}$, the next hour of the frequency time series is predicted by searching the set $\mathcal F$ for the most similar patterns to this initial pattern. We implement this search by computing the Euclidean distance between the initial pattern and all patterns in $\mathcal F$: 
\begin{equation} 
\textrm {d}(\mathbf F_{n}) = \| \mathbf F_{n} - \mathbf F_{0} \|. \label{euclidean}
\end{equation}
The patterns in $\mathcal F$ are then sorted based on these distances in increasing order, thus resulting in a new set of patterns. This set only contains the $k$ most similar patterns to $\mathbf F_{0}$ given as
\begin{equation} 
\mathcal N_{k} = \{ n_{1}, n_{2},\ldots, n_{k} | \mathbf F_{n_{j}} \in \mathcal F \}.
\end{equation} %
The final prediction $f_{wnn}(t_{0} + \Delta t)$ with time steps $\Delta t \in$ \{1s, 2s, 3s, ..., 3600s\} is performed by a weighted average of the $k$ nearest neighbours' next hour trajectories: 
\begin{equation} 
f_{wnn}(t_{0} + \Delta t) = \frac {1}{\sum _{j=1}^{k} \alpha _{j}} \sum _{j=1}^{k} \alpha _{j} f(t_{0} - n_{j}\gamma \tau + \Delta t). 
\label{eq:prediction}
\end{equation}
Note that by the construction of $ \mathcal F $, we restrict our search for frequency patterns to the same hour, e.g. we predict the frequency at 9pm only using training data from 9pm as well, see also \cite{Benjamin,github}.
The weighting $\alpha$ decreases with greater distance between the patterns, since close neighbours should have a stronger influence on the prediction than more distant ones:
\begin{equation} 
\alpha _{j} = \frac {\textrm {d}(\mathbf F_{n_{k}}) - \textrm {d}(\mathbf F_{n_{j}})}{\textrm {d}(\mathbf F_{n_{k}})-\textrm {d}(\mathbf F_{n_{1}})}. \label{weighting}
\end{equation}
The number of nearest neighbours $k$ is a hyperparameter of the WNN predictor, which has to be optimised. Within this paper, we utilize the $adaptive$-$k$ approach \cite{Benjamin}, which optimises $k$ for every time step as follows:  The optimal $k$-value is given by the lowest $\textrm{MSE}_{\Delta t}(f_{wnn})$ of the predictor for each $\Delta t$, i.e. we obtain an $adaptive$-$k$ vector with 3600 values. The $adaptive$-$k$ is smoothed with a sliding window of one minute to avoid sudden jumps and noise for any time steps, see \cite{github} for details. 

As a benchmark, we compare the WNN performance with the daily profile, which effectively is an unweighted $k=\infty$ predictor (including all patterns) defined as
\begin{equation} 
f_{dp}(t_{0} + \Delta t) = \frac {1}{|\mathcal F|} \sum _{n \in \mathcal F} f(t_{0} - n\gamma \tau + \Delta t).
\label{eq:daily_profile}
\end{equation}

\subsection{Performance of the WNN predictor}\label{sec:perform_predcitor}

Let us now investigate whether the WNN predictor can outperform common benchmarks on islands, as it did in continental regions \cite{Benjamin}. We consider three prediction models: first: a naive and constant 50Hz predictor, as this is the typical mean value for each hour, second: the daily profile, which takes all patterns into account equally and third: the WNN, which selects fewer patterns and weights them to obtain a forecast. For all predictors, we use about 70\% of the data for training (i.e. data to select patterns from or data to compute the daily profile), 15\% as validation (to determine the optimal number of nearest neighbours $k$) and 15\% as a test (to evaluate the performance), see the code \cite{github} for details.

\begin{figure}[h]
\centering
\includegraphics[width=13.5cm]{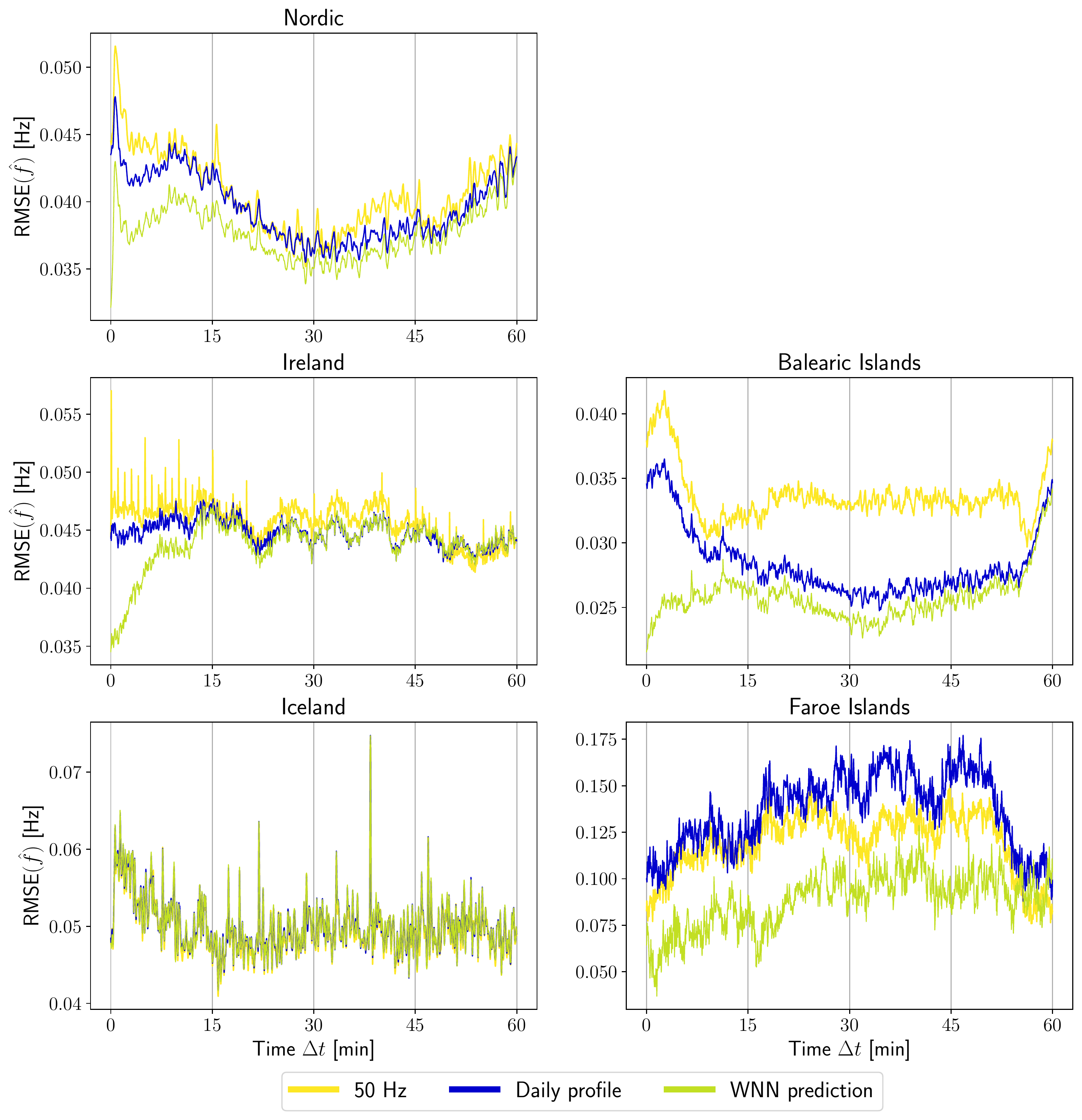}
\caption{For three out of four islands, the WNN predictor outperforms the two benchmark models. 
    We plot the Root Mean Square Error (RMSE) of the WNN prediction model versus the daily profile and a constant 50Hz prediction. In all regions except Iceland, the WNN predictor outperforms the null models, especially the first 15 minutes of the hour. The Balearic Islands have the lowest absolute error, while the Faroe Islands have the highest. The y-axis varies between the subplots due to the varying error between the regions.} \label{fig:performance}
\end{figure}

The WNN predictor typically outperforms these benchmarks even when little data are available, see Fig.~\ref{fig:performance}.
We compute the root-mean-squared error (RMSE) for each predictor for each minute, averaged over all hours. Consistent with the daily profiles, we note that both benchmarks display a large error at the start of an hour, as dispatch actions such as generation or load ramps lead to large deviations in the frequency. Especially at the beginning of an hour, the WNN predictor is more precise than the benchmarks, leveraging its specific information from the previous hour. The further away the prediction is evaluated from the start of the hour, the less specific information is contained in the previous patterns and the WNN approaches the daily profile behaviour, as also seen for continental regions \cite{Benjamin}. There are two more interesting observations: First, even with only 6 days of training and test data available, the WNN outperforms the daily profile and 50Hz predictor in the Faroe Islands. Second, all predictors fare approximately equal for Iceland, where the grid frequency time series shows almost no characteristic patterns but large stochastic deviations. 
\revise{Intuitively, power grids should be easier predictable, the more their trajectories follow deterministic patterns, compared to stochastic influences. Let us discuss this in terms of increment statistics: Large and particularly non-Gaussian increments signal many erratic (random) jumps, while smaller increments signal more regular dynamics. Indeed, we observed for Iceland clear non-Gaussian increments in Fig.~\ref{fig:increments}. Meanwhile, we also observe non-Gaussian increments for Ireland, for which both the daily profile and WNN predictor yield lower forecasting errors. Hence, increment statistics might provide some possible explanations why certain areas are harder to predict than others but further comparisons and metrics will have to be explored in the future.}

In many situations, we will not have access to years or months of data, due to a large variety of potential reasons: Measurement time is limited, large parts of the data are corrupted, the specific grid is very new, either because the grid is newly built or major changes were made, such as the expansion of a synchronous area \cite{bottcher2022initial} or a change in regulations. 
Regardless of what limits the access to data, we need to understand how well the WNN predicts the frequency with shorter periods of training data available.  

\begin{figure}[h]
\centering
\includegraphics[width=10cm]{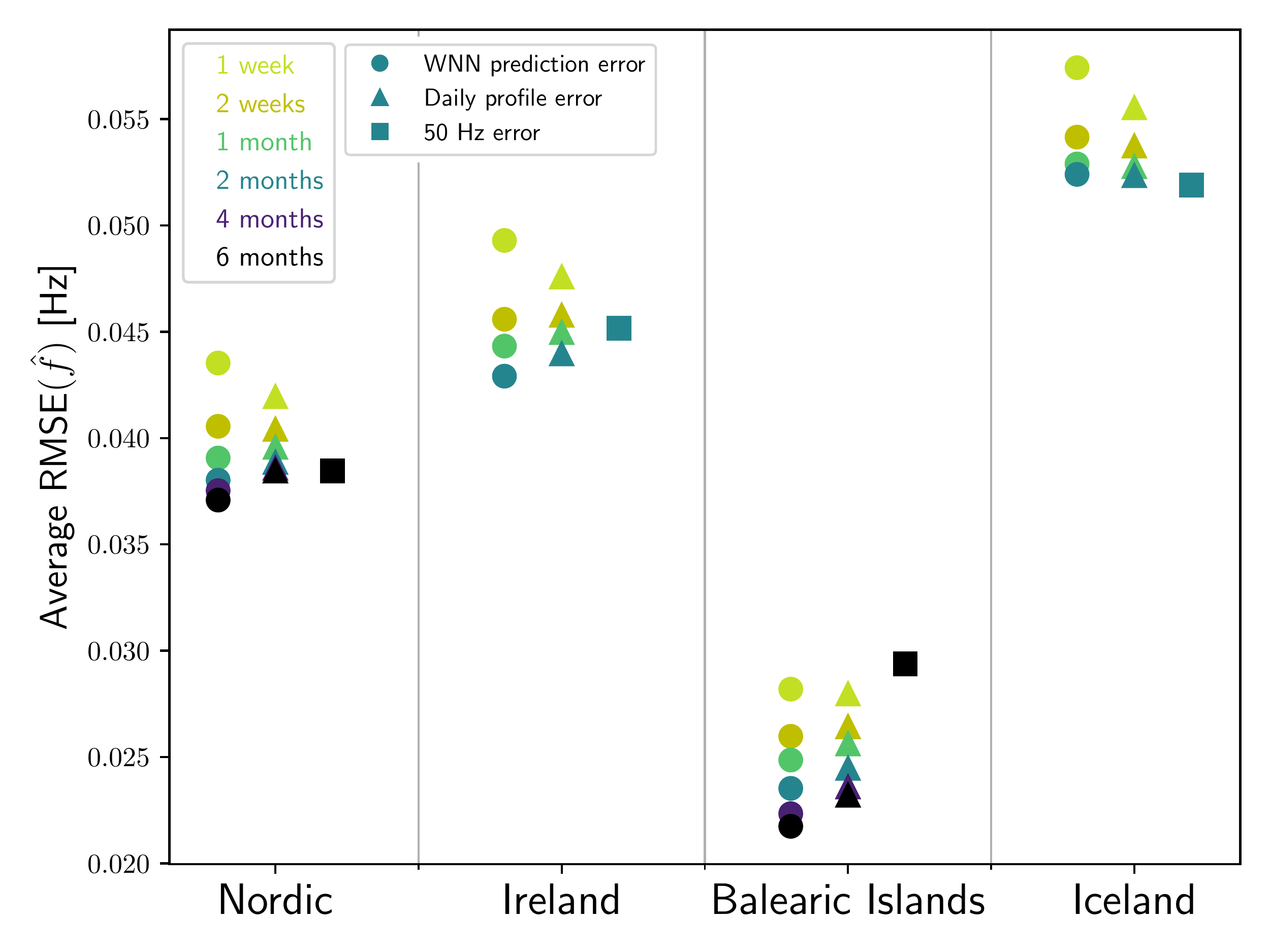}
\caption{With sufficient data, the WNN predictor outperforms the daily profile at three out of four islands. 
We plot the average root mean square error for the WNN predictor and the two null models for different time intervals. As the data size increases, the WNN predictor is the best model for all regions except Iceland, where the constant 50 Hz prediction performs best. The Faroe Islands are not included in the analysis as the region's frequency time series is shorter than the minimum interval of one week.} \label{fig:size_dependence_performance}
\end{figure}

Interestingly, there is no general threshold of how many weeks of data are necessary for the  WNN to outperform the daily profile, see Fig.~\ref{fig:size_dependence_performance}. 
There, we systematically investigate the prediction quality as a function of the length of available data by computing the RMSE for the constant 50Hz prediction, the daily profile and the WNN. The latter two predictors are based on data ranging from one week up to 6 months (where available) and always utilised an 80-20 split for training and validation, hence reporting the validation RMSE. Again, the Faroe Islands are excluded as less than one week of data are available. 
We typically note that for very little data, the constant prediction is the best, followed by the daily profile and the WNN coming in last. If measurement times are too short, the unspecific, average prediction is superior and the WNN restricts itself to finding patterns in its very limited training set, leading to a bad performance. Vice versa, if sufficient data are available, the WNN correctly identifies characteristic patterns and is capable to make the best predictions. 

Notably, this transition from "unspecific information being best" to "specific historic patterns being best" depends on the power grid at hand: The Balearic islands with their very deterministic frequency profile, only require 2 weeks for the WNN to outperform the daily profile, compared to about 4 weeks necessary in Ireland, while in Iceland even with 2 months of data, the constant 50 Hertz prediction is still better than daily profile or the WNN predictor. 

\subsection{Additional features}\label{sec:additonal_feat}

So far, we have only used historic frequency information to forecast frequency time series. However, there are  an increasing amount of energy system data available, so we ask: How can additional information, such as the generation of coal power plants or volatile renewable generation (wind and solar power), be integrated into the frequency prediction, and how do they influence the performance of the WNN predictor? 

To answer this question, we utilise additional information on the Balearic energy system, obtained from the transmission system operator of Spain, Red El\'ectrica de Espa\~na, and their real-time electricity demand and generation tracker \cite{real_time}. It seems intuitive that (frequency) patterns of the power system will differ based on the total load or the mix of generation types. 
Mathematically, we extend the WNN predictor as follows: Let $a_\text{origin}$ be the original additional time series. To compensate for the large deviations in the typical values of $a_\text{origin}$, we carry out a min-max scaling:
\begin{equation}
    a = \frac{a_\text{origin} - a_\text{min}}{a_\text{max} - a_\text{min}},
\end{equation}
where $a$ is the transformed data point and $a_\text{max}$ and $a_\text{min}$ are the maximum and minimum values of the original time series.
Next, the additional feature is split into non-overlapping patterns $\mathbf A_{n}$, similar to $\mathbf F_{n}$ from eq.~\eqref{F}:
\begin{equation}
\mathbf A_{n} = \begin{pmatrix}
a(t_{0} - (n+1)\eta \lambda) \\ a(t_{0} - (n+1)\eta \lambda + \lambda) \\ a(t_{0} - (n+1)\eta \lambda + 2 \lambda) \\ {\dots }\\ a(t_{0} - n\eta \lambda - \lambda) \end{pmatrix},
\end{equation}
where $\eta$ is the number of additional data points in each pattern, and $\lambda$ is the time resolution. $\eta \lambda$ is thus the prediction window size and must be the same size as the frequency patterns to represent the same specific time window of 3600 seconds, matching $\mathbf F_{n}$. Hence, $\eta \lambda \stackrel{!}{=} \gamma \tau$. When the vectors $\mathbf A_{n}$ are concatenated with $\mathbf F_{n}$, the result is new vectors given as:
\begin{equation}
\mathbf G_{n} = \begin{pmatrix} \mathbf F_{n} \\ \mathbf A_{n} \end{pmatrix}.
\end{equation}
As the extended model is also limited to searching through patterns from the same time of the day, $\mathcal G$ is a new set similar to $\mathcal F$ from (\ref{time_sensitive}). To now choose the nearest neighbours in terms of distance, the previous distance computation eq.~ \eqref{euclidean} is extended to
\begin{equation}
\textrm {d}(\mathbf G_{n}) = \| \mathbf F_{n} - \mathbf F_{0} \|+ \beta \cdot \| \mathbf A_{n} - \mathbf A_{0} \|, 
\end{equation}
where $\| \cdot \|$ denotes the Euclidean distance, and $\beta$ is a tunable weight to adjust the influence of the additional feature. With new distances, the weighting $\alpha$ from (\ref{weighting}) is now computed as
\begin{equation} 
\alpha _{j} = \frac {\textrm {d}(\mathbf G_{n_{k}}) - \textrm {d}(\mathbf G_{n_{j}})}{\textrm {d}(\mathbf G_{n_{k}})-\textrm {d}(\mathbf G_{n_{1}})}.
\end{equation}
The weights are then used as in eq.~\eqref{eq:prediction} to forecast the frequency time series.

\begin{figure}[h]
\centering
\includegraphics[width=10cm]{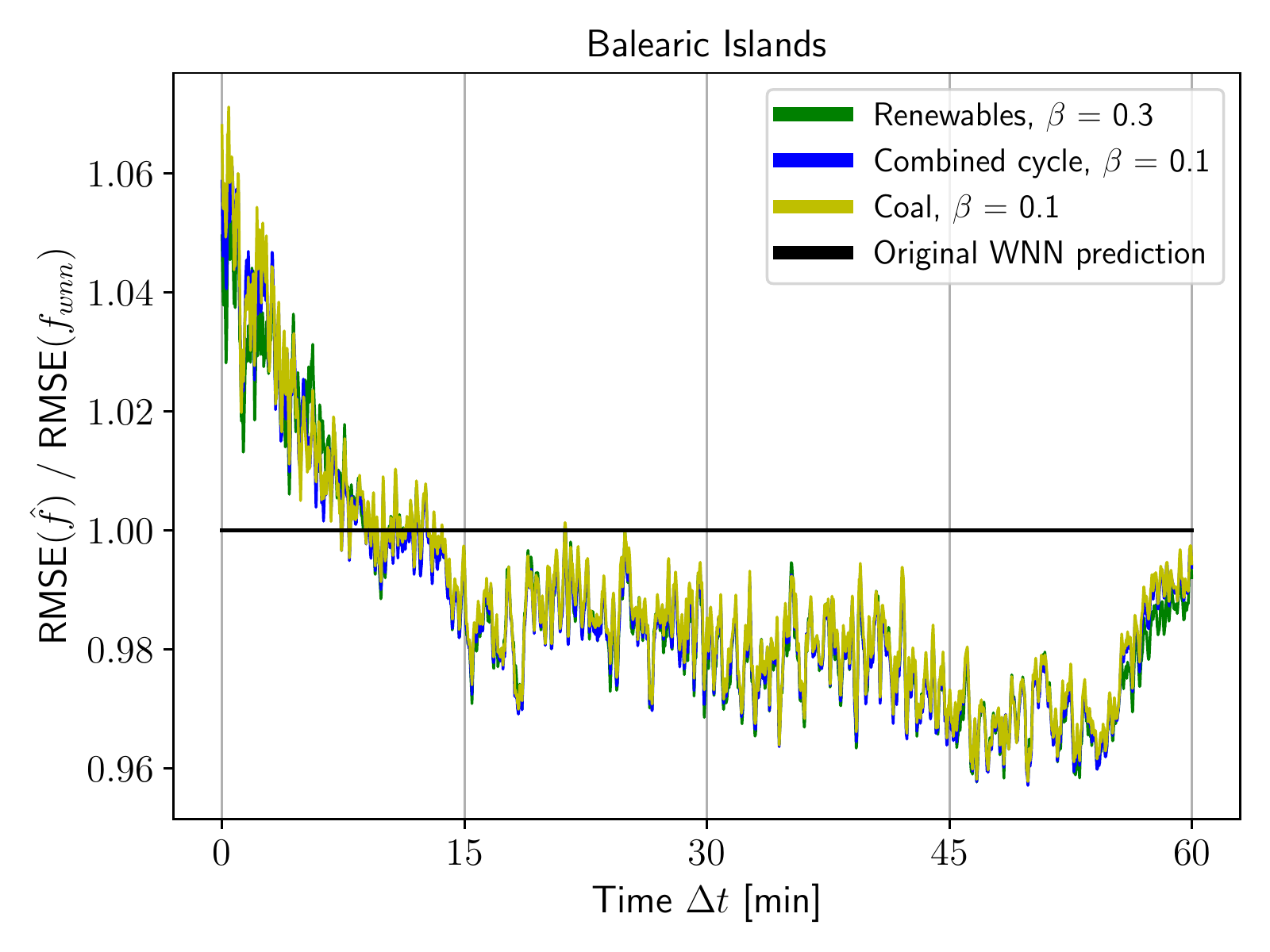}
\caption[Additional feature performance on the Balearic Islands]{With additional features included in the WNN predictor for the Balearic Islands, the prediction performance is increased. Presented are the three best-performing features, all improving the performance by an average of 1-1.5\%. Across the full hour, the first 15 min got worse, but the last 45 min significantly improved.} \label{fig:additional_feature_performance}
\end{figure}

Additional information might improve performance slightly for certain time periods, see Fig.~\ref{fig:additional_feature_performance}.
To obtain these results, we optimized the hyperparameter $\beta$ via a grid search $\beta \in {0.1, 0.2, ..., 1.5}$
Including the information about actual renewable generation accounted for the best performance improvement, closely followed by improvements by including data from combined cycle or coal \cite{red} generation, all of which lead to a reduction in RMSE of more than 1\%. 
Interestingly, the additional features decreased the performance during the first 15 minutes, while it improved after that. Without additional features, the predicted trajectory has its best performance at the start of an hour and then deteriorates since the information of the previous hour is slowly being invalidated by additional, unforeseen events, see also \cite{Benjamin}. 
The additional feature balances this effect: In the first few minutes, the additional feature dilutes the specific information, leading to a larger RMSE, while in the later parts of the hour, the additional information is more useful.
Further research will be necessary to clarify the exact effects of additional features included in the WNN predictor, especially when applying it to other islands, such as Iceland, for which unfortunately no public data on demand and generation were available when writing this article.
Moreover, in our current implementation, we use the actual generation for the hour for which we forecast the frequency. Thereby, we mix past and future information. Future research could rely on forecasts or scheduled generation, which are not always publicly available but might be available to system operators wishing to implement the WNN predictor.

\section{Discussion and Conclusion}\label{sec:discssion}

The power grid frequency in islands may be predicted via a weighted-nearest neighbour (WNN) predictor \cite{Benjamin}.
Compared to the continental areas considered \cite{Benjamin} and the Nordic area analysed in the current article, the four investigated islands displayed more extreme properties:
The daily profile in the Balearic islands is the most pronounced and regular behaviour we have observed so far, while the frequency dynamics in Iceland show no characteristic patterns but strong stochastic deviations, making it particularly tricky to predict.
We also note some dependency on the population size and the composition of the power generation and consumption:
the Faroe Islands and Iceland are the hardest to predict. These two islands are the least populated areas and have very special generation and demand patterns, influenced e.g. by large industry in Iceland. 
Meanwhile, the very good predictability of the Balearic Islands might at least  be partially explained by its strong interconnection with the Continental European synchronous area and the predominance of combined cycle production with its fast control capabilities \cite{martinez2021data}.
Overall, we have shown that the WNN predictor outperforms the daily profile even when only limited data are available in three out of the four island grids. 
Furthermore, we have demonstrated how additional time series, such as generation, could be included as additional information when forecasting the frequency. 

\revise{Concluding, we make two contributions in this article: First, we demonstrate with a simple nearest-neighbour predictor that the power grid frequency is predictable using machine learning in islands if sufficient data are available. Second, we provide a tool that could be useful for the planning and operation of islanded grids, let these be natural islands or islanded microgrids \cite{microgrids}. Transmission system operators (TSOs) could use these forecasts to estimate the stability conditions for the upcoming hour and both TSOs and balancing power provider could use forecasts to estimate control needs for future time intervals: Large frequency deviations will require additional control to be activated to stabilize the power system.}
Notably, almost no prior knowledge about the system at hand is necessary nor do we require many assumptions to apply the WNN. Even the difference between weekends and weekdays is implicitly included in the WNN: Given sufficient data, it will select suitable patterns.
Furthermore, the WNN could be used as a diagnostic tool, e.g. to analyse systematic offsets in the frequency or external perturbations and thereby detect anomalies.

Further research is necessary to fully characterise to which extent power grids are predictable and which external factors should be included. Also, further ML methods, beyond nearest-neighbour techniques could be considered \cite{lim2021time}, e.g. to handle any concept drift in long-term forecasting or to provide further monitoring capabilities.

\acknowledgements{This project has received funding from the Helmholtz Association under grant no. VH-NG-1727, the Program “Energy System Design” and the Helmholtz Association’s Initiative and Networking Fund through Helmholtz AI. }


\bibliographystyle{apsrev}
\bibliography{references}
\end{document}